\documentstyle[buckow]{article}

\def\be{\begin{equation}}
\def\ee{\end{equation}}
\def\bea{\begin{eqnarray}}
\def\eea{\end{eqnarray}}

\def\to{\rightarrow}
\def\del{\partial}
\newcommand{\RR}{{{\rm {}_{R}}}}

\begin {document}
{\hbox to\hsize{
\hfill ITFA-2001-02}}
\def\titleline{The Holographic Renormalization Group}
\def\authors{Jan de Boer}
\def\addresses{Institute for Theoretical Physics\\
University of Amsterdam\\
Valckenierstraat 65\\
1018 XE Amsterdam, Netherlands}

\def\abstracttext{In this lecture, we review the derivation of the
holographic renormalization group given in hep-th/9912012. Some extra
background material is included, and various applications are discussed.}

\large
\makefront

\section{Introduction}

Soon after Maldacena proposed the AdS/CFT correspondence \cite{ads1},
it was realized that there is a close relation between the radial
flow in AdS space the renormalization group flow of the dual field theory.
Since then, there has been a considerable effort to understand this 
relation in more detail. In this lecture, we review the approach of
\cite{dbvv}, which uses the Hamilton-Jacobi equations of canonical gravity
to derive the holographic renormalization group in a general setting.
In addition, we include some background material and discuss some
applications, in particular to domain wall solutions and brane world
scenarios. We will not attempt to give a complete review or even
a complete list of papers where the holographic renormalization group
is discussed. Some of the earlier papers include
\cite{extr1,extr2,rg1,extr11,rg2,rg11,nicketal,porratietal,rg3,rg4},
for more recent lists of references see for example \cite{rec1,rec2,rec3}. 
See also the contributions of Behrndt, Herrmann, Theisen, Bakas and
Kalkkinen to these proceedings.

A brief outline is as follows. We will first review the AdS/CFT correspondence,
in particular those features that are important in understanding the
holographic renormalization group flow. Next, we briefly discuss these
renormalization group flows and discuss a prototype flow, namely a
domain wall solution. We then remind the reader of the basic idea
of Hamilton-Jacobi theory in classical mechanics. We subsequently
apply Hamilton-Jacobi theory to five-dimensional gravity coupled
to scalar fields, and explain how to extract from those the Callan-Symanzik
equations of the dual field theory. We discuss some applications 
of this technique to the computation of the conformal anomaly,
the cosmological constant problem and brane world scenarios. 
We conclude with some comments and by mentioning open problems.

\section{A brief review of the AdS/CFT correspondence}

The holographic renormalization group finds its origin in the AdS/CFT
correspondence, and many of its qualitative features have a clear interpretation
in that context. Therefore, we first give a brief review of the relevant
aspects of the AdS/CFT correspondence. For a more detailed review and
many references, see \cite{review}.

The standard example of the AdS/CFT correspondence is the duality between
type IIB string theory compactified on $AdS_5 \times S^5$, and four-dimensional
$N=4$ super Yang-Mills theory \cite{ads1}. This duality arises by considering
D3 branes in type IIB string theory. Taking a low-energy limit of the string theory
with the D3 branes yields $N=4$ super Yang-Mills theory, whereas the same
low-energy limit applied to the dual supergravity description of the D3 branes
leaves us with $AdS_5 \times S^5$. The so-called Poincar\'e patch of
$AdS_5 \times S^5$ is described by the metric
\be \label{metric}
ds^2 = dr^2 + e^{2r/L} \eta_{\mu\nu} dx^{\mu}dx^{\nu} .
\ee
Here, $L$ is the radius of of $AdS_5$, $L=(g^2_{YM} N)^{1/4}$. The 
boundary of $AdS_5$ is the region where $r\rightarrow \infty$, and this
is where the field theory ``lives''. The coordinate $r$ is directly related to the
radial distance to the D3 branes in the original description of D3 branes in
type IIB string theory. Notice that the metric (\ref{metric}) has an invariance
which maps
\be \label{scale}
r \rightarrow r+a, \qquad \eta_{\mu\nu} \rightarrow \eta_{\mu\nu} e^{-2 a/L} .
\ee
Thus, the scale set by the metric $\eta_{\mu\nu}$ has no invariant meaning
in (\ref{metric}). In the end, we want to identify $ \eta_{\mu\nu} dx^{\mu}dx^{\nu}$
with the metric of the space on which the field theory lives. We therefore see
that the field theory must be insensitive to the scale of the metric, in other
words, it must be a conformally invariant field theory. Four dimensional
$N=4$ super Yang-Mills theory has this property, and this is one of the first
consistency checks of the $AdS/CFT$ duality. 

In \cite{ads2,ads3} it was described how one computes correlation functions
of the boundary field theory using the bulk supergravity description of
$AdS_5 \times S^5$. It is crucial for this that there is a one-to-one
correspondence between fields in the bulk and operators on the boundary.
Consider a free massive scalar field, with mass  $m$, in an $AdS_5$
background. A generic solution of the field equation behaves for
$r\to\infty$ as
\be \label{asymp}
\phi(r ) \sim 
\alpha e^{(\Delta-4)r/L} + \beta e^{-\Delta r/L} + \ldots
\ee
where $\Delta$ is related to $m$ via
\be \label{massscale}
\Delta(\Delta-4) = m^2 L^2 .
\ee
The transformation (\ref{scale}) rescales the terms in (\ref{asymp}) in
a simple way. That suggests the $\Delta$ is related to the scaling dimension
of the boundary operator that corresponds to the bulk field $\phi$. In fact,
it turns out that $\Delta$ is precisely the scaling dimension. In unitary theories,
scaling dimensions have to be nonnegative. This implies that the mass $m$
has to satisfy the bound
\be \label{breiten}
m^2 L^2 \geq -4 .
\ee
This bound is known as the Breitenlohner-Freedman bound \cite{bfbound}.
Its interpretation in $AdS$ is that $AdS$ space with massive scalar fields
is only stable if the mass of the scalar field obeys (\ref{breiten}). Thus,  
contrary to the flat space, $AdS$ space can support fields with certain
negative mass squared. 

{}From now on we will put $L=1$. We denote by $\phi^I$ a set of bulk
fields, and by ${\cal O}_I$ the corresponding operators of the boundary
theory. The usual statement that relates boundary correlation functions
to bulk quantities is
\be \label{statement1}
\left< e^{\int \hat{\phi}^I {\cal O}_I } \right>_{\rm CFT} = 
e^{-S[\phi^I]_{\rm sugra}} .
\ee
In this equation, $S[\phi^I]_{\rm sugra}$ refers to the supergravity
action, evaluated on a solution of the equations of motion with boundary
condition
\be \label{bcond1}
\phi(r\rightarrow \infty) \sim \hat{\phi}^I .
\ee
More precisely, the right hand side of (\ref{statement1}) should be the
full string partition function evaluated with the boundary conditions
(\ref{bcond1}). In a saddle point approximation, this reduces to the
classical supergravity action evaluated on a solution of the equations
of motion. The left hand side of (\ref{statement1}) is the generating
functional for correlations functions in the boundary CFT.

However, as we see in (\ref{metric}) and (\ref{asymp}), neither the metric
nor the scalar fields become constant as we approach the boundary
of $AdS$. Therefore, (\ref{statement1}) cannot be quite correct as it
stands. An improved version of (\ref{statement1}) that takes the
boundary behavior into account is
\be \label{statement2}
\left< e^{\int \hat{\phi}^I {\cal O}_I } \right>_{{\rm CFT,metric}
\,\,\hat{g}_{\mu\nu}} = 
e^{-S[\phi^I,ds^2=dr^2+g_{\mu\nu}(x,r)dx^{\mu}dx^{\nu}]_{\rm sugra}},
\ee
where $\phi^I$ and $g_{\mu\nu}$ solve the supergravity equations of motion
and behave for large $r$ as
\be \label{bcond2}
\phi^I(r \rightarrow \infty) \sim e^{r(\Delta-4)}\hat{\phi}^I \qquad
g_{\mu\nu}(r \rightarrow  \infty) \sim e^{2r} \hat{g}_{\mu\nu} .
\ee

We have taken one of the two asymptotic behaviors of $\phi^I$ in
(\ref{bcond2}). It was shown in \cite{bkl} that in the solution
\be 
\phi(r ) \sim
\alpha e^{(\Delta-4)r/L} + \beta e^{-\Delta r/L} + \ldots
\ee
the first part corresponds to a source for the operator ${\cal O}$,
and that the second part corresponds to changing the state of
the boundary theory.
In the new state, the operator ${\cal O}$ acquires an
expectation value.
Examples of different
states of  the boundary theory are different point on the
Coulomb branch, which are determined by giving expectation
values to the scalar fields in the gauge theory. The correlation
functions in (\ref{statement2}) are computed with respect to
the vacuum state of the gauge theory, and therefore we need
the boundary behavior as in (\ref{bcond2}). 

It is important to keep in mind that operators with $\Delta<4$  are
relevant, operators with $\Delta=4$ are marginal and those
with $\Delta>4$ are irrelevant. For each of these types of
operators, the asymptotic behavior of the scalar field is different. 
{}For irrelevant deformations, the fields blow up near the boundary,
and (\ref{statement2}) can only be used to compute correlation
functions of such operators, as the full generating function is ill-defined.
{}For marginal and relevant deformations, the full generating function
does make sense, and in the sequel we will mainly have such deformations
in mind. Even for marginal and relevant deformations, it is not at all
clear that reasonable solutions of the equations of motion do
exist. For the metric, it was shown in \cite{fg} that for metrics very
close to the metric (\ref{metric}), good solutions do exist, but
in general the situation is unclear.

Even though (\ref{statement2}) is an improvement over (\ref{statement1}),
it is not yet a complete prescription for the correlation functions, because
the supergravity action will in general diverge. To deal with this problem,
we proceed very much as in field theory, namely we first introduce a finite cutoff,
subtract divergent terms and then take the cutoff to infinity. As cutoff, we
truncate $AdS$ at $r=r_0$. This has the interpretation of a UV cutoff in
the boundary theory \cite{sw}. The $AdS/CFT$ correspondence now becomes
\be \label{statement3}
\left< e^{\int \hat{\phi}^I {\cal O}_I } \right>_{{\rm CFT,metric}
\,\,\hat{g}_{\mu\nu}} = 
e^{-\lim_{r_0 \rightarrow \infty} \left\{ 
S[\phi^I,ds^2=dr^2+g_{\mu\nu}(x,r)dx^{\mu}dx^{\nu}]_{{\rm sugra},r\leq r_0}
-{\rm singular\,\,\,terms} \right\} } .
\ee
The boundary conditions are still as in (\ref{bcond2}). Notice that the supergravity
action in (\ref{statement3}) is evaluated on a space with a boundary. The
action for supergravity on a space with a boundary also contains an extra
boundary term, proportional to the extrinsic curvature of the boundary, which
is needed to have a good variational problem \cite{bterm}.

The singular terms that appear in (\ref{statement3}) depend on the number of
dimensions. On $AdS_5$, there will be a singular term related to the volume
of $AdS_5$ that grows as $e^{4r_0}$. In addition, there will be a term related
to the curvature of $AdS_5$ that grows as $e^{2r_0}$, and there will be a term
connected to the conformal anomaly of the boundary field theory, which  will
be linear in $r_0$. The detailed form of these divergences will be discussed 
later. 

\section{RG flows in AdS/CFT}

As we mentioned in the introduction,
soon after Maldacena's paper on the AdS/CFT correspondence
\cite{ads1} it was realized and examined in examples that
there is a close relation between the $r$-dependence of supergravity
solutions and the  RG-flow of the boundary theory.
One can already see part of this connection
in the discussion of the previous section. If we perform for a large
value of $r_0$ the following substitutions (inspired by (\ref{scale}))
\be \label{scale2}
r_0 \rightarrow r_0+a, \qquad
\hat{g} \rightarrow e^{-2a} \hat{g},\qquad
\hat{\phi} \rightarrow e^{-a(\Delta-4)} \hat{\phi}
\ee
then the supergravity solution will remain unchanged. There
will be a corresponding asymptotic symmetry for the boundary
theory correlation functions. These will be modified by finite
$r$ effects, and by the conformal anomaly, as the conformal anomaly
is the only singular term that does not have an asymptotic invariance
of the form (\ref{scale2}). In any case, the symmetry (\ref{scale2})
is very reminiscent of renormalization group flow, if we make
the following dictionary
\bea 
r_0 & \leftrightarrow & {\rm cutoff}\,\,\, \Lambda \nonumber \\
\hat{g} & \leftrightarrow & {\rm scale} \,\,\, \mu \nonumber \\
\Delta & \leftrightarrow & {\rm (anomalous)\,\,\ dimension } .
\eea
In addition, the subtraction of the singularities in (\ref{statement3})
is very similar to the renormalization procedure of a quantum field
theory. 

To make this connection very precise is difficult, because we do not
know  to which regularization and renormalization scheme the finite
cutoff $r_0$ corresponds in the boundary theory. It seems unlikely
that it is related to one of the conventional renormalization schemes,
like dimensional regularization. Properties that are insensitive to
the details of the renormalization scheme can be computed
both in the bulk and on the boundary and compared to each
other. An example of such a property is the form of
the renormalization group equations. These
will be derived a bit later after we discussed the Hamilton-Jacobi
equations of gravity.

The circle of ideas appearing here has been around for a while.
A relation between the world-sheet and space-time renormalization
group  was discussed in \cite{bda}. The idea that the dilaton is
related to a scale and to an extra dimension appears in non-critical
string theory and matrix theory \cite{matrixreview}, and also
in attempts to view the QCD string as a non-critical string theory
\cite{Polyakov}.

Before proceeding, we give a prototype example of the kind  of
flows in supergravity that one encounters in the AdS/CFT context.
These flows also go under the name of domain walls, and appear
e.g. in \cite{rg3,wall1,wall2}. The one described below is given in
\cite{rg4}. In general, 
such flows are solutions to the equations
of motion of actions of the form
\be \label{flow1}
S=\int d^4 x dr \sqrt{g} \left(
-\frac{1}{4} R + \frac{1}{2} \partial_{\mu}\phi \partial^{\mu} \phi
-V(\phi) \right).
\ee
We consider solutions of the equations of motion that have 4d
Poincar\'e invariance and that are of the form
\be
ds^2 = e^{2A(r )}(\eta_{ij} dx^i dx^j) - dr^2, \qquad
\phi(x,r) \equiv \phi(r ).
\ee
Substituting this form back into (\ref{flow1}), and dividing
out the volume of the 4d flat space, we are left with a
simple $0+1$ dimensional action for the quantities $A(r ),
\phi(r )$,
\bea 
S & = & \int dr e^{4A} 
\left[ 3 (A'(r ) + \frac{1}{3} W )^2 -
\frac{1}{2} (\phi'(r ) - \frac{1}{2} W'(\phi ) )^2 \right]
\nonumber \\
& & - \int dr \frac{\partial}{\partial r} 
\left[ e^{4A} (2A' + \frac{1}{2} W) \right] 
\label{flow2} .
\eea
Here, $W$ is any solution of the equation
\be \label{aux}
V = \frac{1}{8} W'(\phi)^2 - \frac{1}{3} W(\phi)^2 .
\ee
This is similar to the relation between the scalar potential
$V$ and the superpotential $W$ in supergravity theories,
but here it appears naturally without supersymmetry.
The action (\ref{flow2})
 is written in a form reminiscent of BPS equations, namely
as a sum of squares and a total derivative. The squares do not come
with the same sign, so the action is not quite of BPS form, but
it is still true that putting the squares each equal to zero will yield
solutions to the equations of motion. The full set of equations
of motion of (\ref{flow1}) is actually larger than those obtained
from (\ref{flow2}), and reads
\bea
\phi'' + 4 A' \phi' & = & V'(\phi) \nonumber \\
A'' & = & -\frac{2}{3} \phi'^2  \label{eom} \\
A'^2 & = & -\frac{1}{3} V(\phi) +\frac{1}{6} \phi'^2 \nonumber  .
\eea
One easily verifies that once the squares in (\ref{flow2}) are
put equal to zero, these equations are all satisfied. 
Thus, if we pick a $W$
that solves (\ref{aux}), and choose two initial conditions $A(r_1)$
and $\phi(r_1)$ at $r=r_1$, we can 
integrate the equations $A'=-\frac{1}{3}W(\phi)$ and
$\phi'=\frac{1}{2} W'(\phi)$ from $r_1$ to $r_2$ and
find a solution. If we evaluate (\ref{flow2}) for this solution, 
we find 
\be \label{onshell}
S = \frac{1}{6} \left( e^{4A(r_2)} W(r_2) - e^{4A(r_1)} W(r_1)
\right) .
\ee
The action only depends on the values of $A$ and $W$ at the endpoints.
We will later see how we can rederive this result using the Hamilton-Jacobi
equations.

It is an interesting question whether the procedure given here provides
all possible solutions to the equations of motion. In a supersymmetric
context, the first order equations would be forced upon us by requiring
a supersymmetric solution, but in the purely bosonic case this is no longer
obviously the case. Suppose that we are given some smooth solution of the
equations of motion, with $|\phi'(r )|\neq 0$. Then
we can certainly define a function $W(\phi)$ such that $A'(r )=
-\frac{1}{3} W(\phi(r ))$. Using this definition, the second equation in (\ref{eom}) 
implies that $\phi'=\frac{1}{2} W'(\phi)$, and (\ref{aux}) is a consequence
of the third equation in (\ref{eom}), after which the first equation of
(\ref{eom}) is trivially satisfied. The general solution of the equations of
motion is therefore obtained by gluing together solutions of the
set of first order equations. At the glueing point, the potential $V$
satisfies $V'(\phi)=0$ and has an extremum. Thus, solutions can be
naturally decomposed in pieces that roll between the extrema of $V$,
and each of the pieces is described by some set of first order equations. 
If $W$ is given, and both $\phi(r_1)$ and $\phi(r_2)$ are fixed by e.g.
a potential at the endpoints, then the difference between $r_1$ and $r_2$ is
also fixed. This is because the first order system describing the flow
of $\phi$ will only have a solution with the right boundary conditions
if $r_1-r_2$ is tuned in the right way. This dynamical tuning of
the distance between the endpoints is known as the Goldberger-Wise
mechanism \cite{gw}.

\section{Hamilton-Jacobi theory}

The main tool in the derivation of renormalization group  equations
from 5d gravity is the Hamilton-Jacobi formalism. As a reminder,
we briefly review the formalism for quantum mechanics.

Consider a classical action
\be \label{class}
S = \int_{x(t_1)=x_1}^{x(t_2)=x_2} {\cal L}(q,\dot{q}) dt ,
\ee
and denote by $S[x_1,x_2]$ the value of the action evaluated
along the solution of the equations of motion $q_0(t)$ 
with the prescribed
boundary conditions. We are interested in the behavior of
$S[x_1,x_2]$ as we vary $x_2$. If we replace $x_2$ by
$x_2 + \delta x_2$, the solution of the equations of motion
is replaced by $q_0(t) + \delta q_0(t)$. The variation of
$S[x_1,x_2]$ is given by
\bea
\delta S & = & \int {\cal L}(q_0 + \delta q_0, \dot{q}_0 +
\delta \dot{q}_0) - \int {\cal L}(q_0,\dot{q}_0) \nonumber \\
& = & \int_{t_1}^{t_2} \left( 
\delta q_0 \frac{\partial {\cal L}}{\partial q} + 
\delta  \dot{q}_0 \frac{\partial {\cal L}}{\partial \dot{q}}
\right) dt \nonumber \\
& = & \int_{t_1}^{t_2}\delta q_0\left(
\frac{\partial {\cal L}}{\partial q} - \frac{\partial}{\partial t}
\frac{\partial {\cal L}}{\partial \dot{q}} \right)
+ \left. \delta q_0 
\frac{\partial {\cal L}}{\partial \dot{q}} \right|_{t_1}^{t_2}
\nonumber \\
& = & \delta x_2 \frac{\partial {\cal L}}{\partial \dot{q}} (t_2)
= \delta x_2 p(t_2) .
\label{basiceq}
\eea
In the one but last line, the first term vanishes since
we assumed that $q_0$ was a solution of the classical
equations of motion. 
We see that the derivative of $S[x_1,x_2]$ with respect to
$x_2$ is given by the momentum $p(t_2)$. We can therefore
in the Hamiltonian replace momenta by derivatives of the
action. This then finally leads to the Hamilton-Jacobi equation
\be \label{hjeq}
H(q,\frac{\partial S}{\partial q}) = \frac{\partial S}{\partial t} .
\ee
A  convenient way to think about this is to remember that 
path integrals in quantum mechanics with boundary conditions
actually compute wave functions. Wave functions satisfy 
the Schr\"odinger equation 
\be \label{seq}\left(
\frac{\hbar}{i} \frac{\partial}{\partial t} - H(q,
\frac{\hbar}{i} \frac{\partial}{\partial q}) \right)
\Psi = 0 .
\ee
If  we write
\be \Psi = e^{\frac{i}{\hbar} S} \ee
and expand (\ref{seq}) to lowest order in $\hbar$, we recover
the Hamilton-Jacobi equation (\ref{hjeq}). Indeed, the classical
action is the lowest order approximation to the quantum mechanical
path integral (the WKB approximation).

As an example,  consider the relativistic point particle. The 
Hamilton-Jacobi equation reads
\be  \label{aux2}
\frac{\partial S}{\partial x^{\mu}}
\frac{\partial S}{\partial x_{\mu}} = m^2 c^2.
\ee
The right hand side is not equal to $\partial S/\partial t$, but a constant.
This is a consequence of the reparametrization invariance  of the point
particle action. Something similar happens in string theory, where the
Virasoro constraint $L_0=1$ (which also is a consequence of 
reparametrization invariance) fixes the value of the Hamiltonian as well.

There are several solutions to (\ref{aux2}). For instance, for a classical
trajectory we have
\be S = mc \sqrt{  (x-x_0)_{\mu} (x-x_0)^{\mu} } ,
\ee
but $S=a_{\mu} x^{\mu}$, with $a^{\mu} a_{\mu} = m^2 c^2$ also
solves the Hamilton-Jacobi equation. 
 
The Hamilton-Jacobi formalism also applies to gravity in its
canonical form, the latter being 
known as the ADM formalism \cite{adm}. We will be specifically
interested in the five dimensional case, but everything generalizes
in a straightforward way to other dimensions. We use Euclidean 
signature. We can always write the five-dimensional metric in
the form
\be \label{gauge1}
ds^2 = N^2 dr^2 + g_{\mu\nu}(x,r)(dx^{\mu} + N^{\mu}dr)
(dx^{\nu} + N^{\nu} dr)
\ee
where $N$ is known as the lapse function and $N^{\mu}$ as the shift
function. The role of the coordinate $r$ will be similar to that
of a time coordinate, and it is the direct generalization of the radial
coordinate of $AdS_5$. Instead of flows in time, this Hamilton-Jacobi
formalism will describe flows in the radial direction  $r$, which
we expect to be the flow that corresponds to renormalization group
flow.  Locally, we can use the diffeomorphism invariance
to choose $N=1$ and $N^{\mu}=0$, and we will work in this
gauge from now on. However, we must still impose
the $N$ and $N^{\mu}$ equations of motion, and these give rise
to the Hamilton constraint and the diffeomorphism constraint respectively.
The Hamilton constraint will provide us with the Hamilton-Jacobi
equation. 

The system we will consider is five-dimensional gravity, minimally
coupled to a set of scalars $\phi^I$, with potential $V(\phi)$,
and kinetic term with $\frac{1}{2} \partial \phi^I G_{IJ}(\phi)
\partial \phi^J$. The variables of the theory are therefore the
metric $g^{\mu\nu}$ and the scalars $\phi^I$. The corresponding
canonical momenta will be denoted by $\pi_{\mu\nu}$ and
$\pi_I$. The momenta can, as above, be related to derivatives
of the classical action $S$ with respect to the boundary conditions.
In this case, we choose as our boundary a slice at fixed values of $r$,
and $S[\phi,g]$ will be the classical supergravity action, evaluated
on a solution of the equations of motion, with boundary conditions
$g$ and $\phi$ at the boundary. As we mentioned previously, the
classical action for gravity on a manifold with a boundary has
some extra boundary terms in order to have a good variational problem,
and these boundary terms have to be taken into account in the
evaluation of the classical action $S$. 

The Hamilton constraint takes the form 
\be \label{ham1}
{\cal H}
=
\Bigl(\pi^{\mu\nu}\pi_{\mu\nu}\! -{{1\over 3}} \pi^\mu_\mu \pi^\nu_\nu\Bigr)
+ {{1\over 2}} \pi_{I}G^{IJ}(\phi)\pi_{J}
+ {\cal L}(\phi,g)=0,
\ee
where ${\cal L}$ denotes
the local lagrangian density 
\be
{\cal L}(\phi,g) = V(\phi) + R +
\frac{1}{2} \, \partial^\mu\phi^I G_{IJ}(\phi)\, \partial_\mu \phi^J .
\ee
Here we have chosen to work in
the 5-dimensional Einstein frame, so that we can use 5-d Planck units
with Newton constant equal to one.

The 5-dimensional supergravity
equations of motion are implied by the standard Hamilton equations
coming from ${\cal H}$,
supplemented with the diffeomorphism constraint 
\be
\label{conv}
\nabla^\mu\pi_{\mu\nu}+\pi_I\nabla_\nu\phi^I=0,
\ee
as well as the Hamilton
constraint (\ref{ham1}).

Our goal will be to express the constraints (\ref{ham1}) and
(\ref{conv}) in terms of $S[\phi,g]$.
The momenta are expressed in terms of the action $S[\phi,g]$ via
\be \label{defmom}
\pi_{I}={1\over {\sqrt{g}}}{\delta S\over \delta \phi^I}, 
\qquad \pi_{\mu\nu}
=\frac {1}{\sqrt{g}}\frac{\delta S}{\delta g^{\mu\nu}}. 
\ee 
Let us insert these relations into the constraints (\ref{conv}) and 
(\ref{ham1}). Equation (\ref{conv}) gives
\be
\nabla^\mu{\delta S\over\delta g^{\mu\nu}}+\nabla_\nu\phi^I{\delta S\over\delta\phi^I}=0.
\ee
This constraint is easily satisfied: it simply 
means that the effective action $S$ is invariant under 4-d coordinate 
transformations. Equation (\ref{ham1}) takes the form
\be
\label{ham}
\frac {1}{\sqrt{g}}\left({{1\over 3}}
\Bigl(g^{\mu\nu}\frac{\delta S}{\delta g^{\mu\nu}}\Bigr)^2-
g^{\mu\lambda} g^{\nu\rho}\frac{\delta S}{\delta g^{\mu\nu}}
\frac{\delta S}{\delta g^{\lambda\rho}} 
-\frac {1}{2} G^{IJ}{\delta S\over \delta \phi^I}
{\delta S\over \delta \phi^J}\right)
= \sqrt{g} {\cal L}(\phi,g).
\ee
This is the Hamilton-Jacobi constraint 
of gravity coupled to scalars and it will play a central role in the remainder.  
It is important to realize that it is not a 
constraint on the {\it value} of the variations of the action $S$ nor an 
equation of motion of the 4-dimensional fields: instead one must read it 
as a functional differential equation that determines the 
functional form of the classical action $S$. 

Given a solution to the Hamilton-Jacobi equation, we can 
compute the radial derivatives of $\phi^I$ and $g_{\mu\nu}$
from the standard Hamiltonian equation of motion $\dot{q}=
\partial H /\partial p$. Using (\ref{ham1}) we obtain
\be \label{radder}
\frac{\partial \phi^I}{\partial r}=G^{IJ}\pi_J ,\qquad
\frac{\partial {g}_{\mu\nu}}{\partial r}= -2\pi_{\mu\nu}+ {2\over 3}
\pi^\lambda_\lambda \, g_{\mu\nu},
\ee
which together with (\ref{defmom}) expresses the radial
derivatives purely in terms of $S$. 

\section{The holographic renormalization group}

In the discussion of the AdS/CFT correspondence, we saw that
in order to compute correlation functions, we had to compute
the supergravity action at a finite value of $r_0$, and
subtract divergences that arise as $r_0\rightarrow \infty$. These
divergent terms are local expressions (exactly  as it happens in
renormalizable field theories), and the effective action of
the gauge theory, which we will denote by $\Gamma$, differs
from $S$ by these local, divergent terms. This motivates the
decomposition
\be
\label{deco}
S[\, \phi\, ,\, g\, ] \, = \, S_{\rm loc}\, [\, \phi\, ,\, g\,] 
\; + \; \Gamma\, [\, \phi\, ,\, g\,] .
\ee
{}For the local part we will take
\be
S_{\rm loc}\, [\, \phi\, ,\, g\, ] \, =  \;
\int \!\! \sqrt{g}  \, \Bigl(U(\phi)  +   
\Phi(\phi) R \, +  \frac{1}{2}\partial^\mu\phi^I M_{IJ}(\phi)\, 
\partial_\mu \phi^J \Bigr),
\ee
where $U$, $\Phi$ and $M_{IJ}$  are local functions of the couplings, while
$\Gamma[\phi,g]$ contains all higher derivative and non-local 
terms. If $S$ would have a good low-energy local derivative expansion,
we could simply define $S_{\rm loc}$ as the first two terms in
this expansion. However, in general $S$ will be nonlocal. To define
(\ref{deco}) properly, we need to use a scaling procedure. 
Naively, one could look at how things scale under rescalings of
the metric, but again, because $S$ is nonlocal, this will not
work properly. The right thing to do is to assign degree $+2$
to $g_{\mu\nu}$, and degree $\Delta_I-4$ to $\phi^I$. This is 
motivated by (\ref{scale2}). We now define $S_{\rm loc}$ to be
a local term that contains at least the complete part of $S$
that has degree larger than zero. 
There is an ambiguity in
(\ref{deco}) that amounts to moving terms with zero and negative
degree between $S_{\rm loc}$ and $\Gamma$. However, the Hamilton-Jacobi
theory will provide us with a nice way to choose an appropriate
set of local terms in $S_{\rm loc}$, and the terms with 
positive degree are fixed unambiguously in this way.
The whole procedure only makes sense for marginal and relevant
perturbations. For irrelevant perturbations there can
be terms of arbitrary positive degree. 

The crucial point in doing the decomposition (\ref{deco}) 
is that the part of $S$ of positive degree
is purely local.
It is not a priori clear that this will be so in full generality. However,
on spaces that are asymptotically $AdS$, once one is
sufficiently close to the boundary, one can explicitly study solutions
to the equations of motion and verify that this is indeed the case.
Thus, for the applications that we have in mind, (\ref{deco}) is
certainly true, but it would be interesting to have a more complete
understanding of this issue. 

In dimensions different from five, one can make expansions similar
to (\ref{deco}). In $d$ dimensions, one should put all local terms
with up to $d-2$ derivatives in $S_{\rm loc}$. The equations that
follow will get more complicated as the number of dimensions increases,
but conceptually the strategy will be identical to the one we will
follow here.

If the space is asymptotically AdS, then when we approach the boundary,
the local potential term will
contain quartic divergences, while the other local terms are 
quadratically divergent. In fact, even $\Gamma$ contains divergences of a 
logarithmic type. 
We further note that the local action $S_{\rm loc}$ is
similar in structure as the lagrangian term ${\cal L}$ in the Hamilton 
constraint (\ref{ham}). In fact, as we will see, the different
terms in $S_{\rm loc}$ have a direct relation with the corresponding 
terms in ${\cal L}$. 

The idea of the following calculations will be to insert the expansion
(\ref{deco}) into the Hamilton constraint (\ref{ham1}), combine the
contributions on the left hand side that have the scaling degree
as the terms on the right hand side, and require them to cancel. 
{}For this purpose, we define
\bea
S^{(0)}_{\rm loc}
\, [\, \phi\, ,\, g\, ] &  =  & 
\int \!\! \sqrt{g}  \, U(\phi)  \nonumber \\
S^{(2)}_{\rm loc}\, [\, \phi\, ,\, g\, ] & =  &
\int \!\! \sqrt{g}  \, \Bigl(   
\Phi(\phi) R \, +  \frac{1}{2}\partial^\mu\phi^I M_{IJ}(\phi)\, 
\partial_\mu \phi^J \Bigr)
\nonumber \\
{\cal L}^{(0)}(\phi,g) & = & \sqrt{g} V(\phi) 
\nonumber \\
{\cal L}^{(2)}(\phi,g) & = &  \sqrt{g} ( R +
\frac{1}{2} \, \partial^\mu\phi^I G_{IJ}(\phi)\, \partial_\mu \phi^J ) .
\eea
The Hamilton-Jacobi equations can be written in the form
\be \label{aux3}
\{ S,S \} + {\cal L}^{(0)} + {\cal L}^{(2)} = 0
\ee
where we define the bracket $\{S,S\}$ through
\be
\{S,S\} = 
\frac {1}{\sqrt{g}}\left({{1\over 3}}
\Bigl(g^{\mu\nu}\frac{\delta S}{\delta g^{\mu\nu}}\Bigr)^2-
g^{\mu\lambda} g^{\nu\rho}\frac{\delta S}{\delta g^{\mu\nu}}
\frac{\delta S}{\delta g^{\lambda\rho}} 
-\frac {1}{2} G^{IJ}{\delta S\over \delta \phi^I}
{\delta S\over \delta \phi^J}\right).
\ee
If we expand (\ref{aux3}) in scaling degree, as explained above,
we get various equations, the first three of which we list below:
\bea
\{ S^{(0)}_{\rm loc}, S^{(0)}_{\rm loc}\} & = & {\cal L}^{(0)} \label{eq1}
\\
2 \{ S^{(0)}_{\rm loc}, S^{(2)}_{\rm loc} \} & = & 
{\cal L}^{(2)} \label{eq2} \\
2\{S^{(4)}_{\rm loc}, \Gamma \} + 
\{ S^{(2)}_{\rm loc}, S^{(2)}_{\rm loc} \} & = & 0 .
\label{eq3}
\eea

These equations are not strictly speaking necessary, since 
e.g. $S^{(0)}$ can also contain terms of scaling weight $\leq 0$.
This corresponds to an ambiguity in (\ref{deco}), 
because such terms can be shifted between
$S^{(0)}$ and $\Gamma$. Our point of view will be to 
impose the equations (\ref{eq1}), (\ref{eq2}) and (\ref{eq3})
also for those terms, and to thereby fix most of the 
ambiguity in the decomposition (\ref{deco}). The results
we get show that this is a very natural thing to do.

Let us first look at the first of these equations, (\ref{eq1}).
Collecting the various terms yields the identity
\be
\label{fst}
V =  \frac{1}{3}
U^2 - \frac{1}{2}\del_I U \, G^{IJ} \del_J U. 
\ee
Up to some trivial rescalings, this is the same relation we found in
(\ref{aux}). Thus, from this simple bosonic analysis we recover
the usual relation between the scalar potential and super potential
of supergravity. The first order equations that solved (\ref{eom})
are also easily obtained. They  read
\be \label{linflow}
\dot{\phi}^I = G^{IJ} \partial_J U , \qquad 
\dot{g}_{\mu\nu} = -\frac{1}{3}U(\phi) g_{\mu\nu}, 
\ee
and are obtained from (\ref{radder}) in the case where
we consider Poincar\'e invariant solutions of the supergravity
system only. {}For such solutions, $S\equiv S_{\rm loc}^{(0)}$
and (\ref{eq1}) is the only nontrivial equation. We thus completely
recover all the information about the simple flow disussed around
(\ref{eom}) directly from the Hamilton-Jacobi equation.

The flow equations (\ref{linflow}) can be solved with the ansatz
\be
g_{\mu\nu}\, =\, a^2\, \widehat{g}_{\mu\nu},
\ee
where $\widehat{g}_{\mu\nu}$ is independent of $r$ and the prefactor $a$ 
satisfies 
\be
\label{adot}
\dot{a}\, =\, -\frac{1}{6}\, U(\phi)\,a.
\ee
Since the parameter $a$ in fact determines 
the physical scale (as is also clear from
our previous discussion the AdS/CFT
correspondence), we now replace  the 
$r$ derivatives in the flow equations
by derivatives with respect to $a$, by using 
the relation (\ref{adot}). In this way we obtain
\be
a{d\over da}{\phi}^I \, =\, \beta^I(\phi)
\ee
where the beta-functions are defined by
\be
\beta^I(\phi)=-{6\over U (\phi)}G^{IJ}(\phi)\partial_J U (\phi).
\ee
These beta-functions contain the scale dependence of the fields
$\phi^I$. Near the AdS boundary 
\be
\beta^I(\phi) \sim (\Delta_I-4)\phi^I + \ldots .
\ee
As we will see shortly, these $\beta$-functions indeed play
the role of $\beta$-functions in the renormalization group
equations.  

Notice that the solution of $U$ to (\ref{fst}) is not unique.
{}For supersymmetric flows, we should identify $U$ with the
superpotential, but in general this relation need not hold.
To get an idea about the ambiguities in $U$, we can do
perturbation theory around a critical point of $V$, which
is also a critical point of $U$. 
{}For simplicity we go to a basis in which the metric $G_{IJ}$ is given by 
$\delta_{IJ}$, so that we can write the relation as
\be
\label{A1}
{1\over 3}U^2-{1\over 2}(\partial_I U)^2=V
\ee 
One might wonder whether any potential $V$ can be written in this form.
Our derivation of this relation did not assume any special properties of $V$,
except the existence of a classical solution that can be extended from the
boundary to the interior. This of course puts restrictions on the
potential $V$, and apparently forces it to be of the supersymmetric form.
Let us expand the 5-d potential in powers of $\phi^I$ as
\be
\label{A2}
V\,= 12-{1\over 2} 
m^2_{I}\phi^I\phi^I+g_{IJK}\phi^I\phi^J\phi^K\ldots
\ee
where we used the freedom to shift the fields to remove a possible linear term
in $\phi^I$.  To obtain a solution to (\ref{A1}) for the 4-d potential,
we first try a similar expansion.
\be
\label{A3}
U\,= 6+{1\over 2}\lambda_{I}\phi^I\phi^I+
\lambda_{IJK}\phi^I\phi^J\phi^K
\ee
Here we already fixed the constant term so that it matches with that 
of $V$. The beta-functions derived from this potential are
\be
\beta^I = -(4-\Delta_I) \phi^I - c^I{}_{JK} \phi^J \phi^K
\ee
where
\be
\label{A4}
\Delta_I = \, 4-{\lambda_I} \qquad \qquad \ 
c_{IJK} = \, {3} \, \lambda_{IJK}
\ee 
are the scaling dimensions and operator product coefficients of the 
operators $O_I$ corresponding to the couplings $\phi^I$.
Inserting both expansions (\ref{A2}) and (\ref{A3})
in (\ref{A1}) then gives the relation
\be
\lambda_I^2-4 \lambda_I=m^2_I
\ee
which upon inserting the relation (\ref{A4}) is recognized as the
standard relation (\ref{massscale})
between the scale dimensions $\Delta_I$ of the 4-d couplings
and the corresponding masses $m_I$ of the 5-d fields \cite{ads2}
\cite{ads3}.  Notice that this relation implies that the 5-d
potential must in fact satisfy the unitarity bound $m^2_I \geq
-4$. If this inequality is violated, there are no
bounded solutions that extend all the way to the asymptotic boundary.
Interestingly, this is precisely the Breitenlohner-Freedman bound
(\ref{breiten}) for stability in AdS space. Thus, the existence of
a perturbative solution immediately implies that the masses of
the fields have to be consistent with an AdS solution. Existence
of solutions is also known to be related to the stability of
domain wall solutions of supergravity \cite{rg3,wall1,wall2}.

By looking at the next order in the $\phi^I$ expansion we obtain the
relation
\be
(\lambda_I+\lambda_J+\lambda_K-4\lambda) \lambda_{IJK}= -g_{IJK},
\ee
which via (\ref{A4}) expresses the operator product coefficients
$c_{IJK}$ in terms of the cubic term in $V$. Note that this expression 
degenerates in the case that $\Delta_I + \Delta_J + \Delta_k = 8$.
The interpretation of this will become clear when we look at deformations
of $U$ that preserve the relation (\ref{A1}) with $V$.

An infinitesimal variation $\delta U$ preserves (\ref{A1}) if it
satisfies the linear relation
\be
4 U \delta U - 6 \partial_I U \partial_I(\delta U) = 0,
\ee 
or
\be
(4+ \beta^I \partial_I ) \delta U = 0.
\ee
These equations tell us that those terms in $U$ that have total dimension $4$ 
are not determined by the Hamilton-Jacobi constraint. It is interesting
to note that these are precisely those terms that remain finite in the 
continuum limit. Therefore, it appears that the Hamilton-Jacobi relation
only constrains the divergent terms of the potential $U$, but not the 
finite part. Precisely these terms survive as finite local terms in the boundary
effective action. 

Besides such continuous ambiguities, there are also discrete ambiguities
in $U$. For example, $\lambda_I$ can be either $\Delta_I$ or $4-\Delta_I$. 
In case the space is asymptotic AdS, the boundary conditions in the IR
will tell us that the right solution is $4-\Delta_I$, but in more
general situations more general $U$ can appear.
We should also point out that some flows to non-susy fixed points can have
$U$ non-analytic at $\phi=0$, in which case the above perturbative expansion does not
apply \cite{sixdexam}.

Another potential worry is that some of the flows under consideration
describe a flow to a Coulomb branch of the gauge theory, i.e. a flow
with $\langle {\cal O} \rangle \neq 0$. To analyze whether one
is really computing correlation functions in the vacuum state or
not requires a careful study of the subleading behavior of the fields
as $r \rightarrow \infty$. 

As an extension of these methods, one can also study the behavior of
correlation functions along flows. This has been done in e.g.
\cite{co1, co2} and yields quite reasonable results. 

We proceed by investigating equation (\ref{eq2}) . This equation
implies
\bea
\beta^K\!\partial_K\Phi &  =&  -2\Phi+{6\over U} \nonumber \\
-2M_{IJ}+{6\over U}G_{IJ} & = & 
-12 \partial_I \partial_J \Phi +
\beta^K\partial_K M_{IJ}\!-\!\beta^K \partial_I M_{KJ}\!-\! 
\beta^K \partial_J M_{IK} \nonumber \\
\beta^I & = & -6 M^{IJ}\partial_J\Phi. \label{secondflow}
\eea
In most studies of the AdS/CFT correspondence, the local terms of the
boundary effective action are considered to be non-universal, since they
diverge as we take the limit $r\to\infty$. Here we see, however, that
they contain very important information about the flow of the couplings 
constants, and even conspire in an interesting way to make up the lagrangian
term ${\cal L}$ of the 5-d gravity theory.  

\section{Callan-Symanzik equation}

As we explained in our discussion of the AdS/CFT correspondence,
the functional $\Gamma$ contains all the information about the
correlation functions of the theory via the identity
\be \label{aux4}
{\large\langle} \, {\cal O}_{I_1}(x_1)\ldots {\cal O}_{I_n}(x_n) \,
{\large\rangle} = 
{1\over{\sqrt {g}}}\,
{\delta \ \ \over \delta\phi^{I_1}(x_1)}\ldots{1\over{\sqrt {g}}}\,
{\delta\ \  \over \delta\phi^{I_n}(x_n)}\, \Gamma\, [\, \phi\, ,\, g\, ]. 
\ee
The Callan-Symanzik equation can be derived, using (\ref{aux4}),
from the last equation we wrote in (\ref{eq3}). That equation implies
\be
\label{loflo}
{1\over \sqrt{g}}
\Bigl( 2 g^{\mu\nu}{\delta\ \over \delta g^{\mu\nu}}-
\beta^I(\phi){\delta \ \over\delta \phi^I}\Bigr)\, \Gamma\, [ \, \phi, \, g \, ]
\; =\ \mbox{4-derivative
terms} 
\ee
In order to derive the Callan-Symanzik equations for expectation
values of local operators we vary this relation with
respect to fields $\phi^{I}$ as in (\ref{aux4}).
 After doing the variations, the 
fields are put to their constant average value given by 
the couplings of the gauge theory. We further take the metric
to be of the form $g_{\mu\nu}=a^2\eta_{\mu\nu}$, where $a$ is
$x^\mu$-independent. The 4-derivative terms will drop out 
after this step, and play no role as long as one considers operators
${\cal O}_I$ at different points in space. Finally, we integrate
the resulting expression over all of space and replace the functional
derivatives by ordinary derivatives by using the definitions
\be -2
\int
g^{\mu\nu}\!{\delta\ \over \delta g^{\mu\nu}}=a{\partial\ \over\partial a},
\qquad\qquad\int {\delta \ \over\delta \phi^I}={\partial\ \over \partial\phi^I}
\ee
In this way one derives after some straightforward algebra from (\ref{loflo})  
the standard form of the Callan-Symanzik equations
\be
\label{CalSym}
\Bigl(a{\partial\ \over \partial a}+\beta^I \partial_I\Bigr){\large\langle}
{\cal O}_{I_1}(x_1)\ldots {\cal O}_{I_n}(x_n){\large\rangle} +
\sum_{i=1}^n\gamma\mbox{\raisebox{-2pt}{${}_{I_i}{}^{\!\! J_i}$}}\,
\langle{\cal O}_{I_1}(x_1)..{\cal O}_{J_i}(x_i).. {\cal O}_{I_n}(x_n)\rangle
\, = \, 0
\ee
where 
\be
\gamma\mbox{\raisebox{-2pt}{${}_I{}^J$}}=\nabla_I\beta^J 
\ee
represent the anomalous scaling 
dimensions of the operators ${\cal O}_I$.
By an appropriate choice of contact term, the ordinary  
$\phi^I$ derivative is turned in a covariant derivative $\nabla_I$ that is 
defined in terms of the metric $G_{IJ}$. This ensure that the whole formalism
stays covariant under field redefinitions. .

The Callan-Symanzik equation (\ref{CalSym})
is derived still with a finite cut-off. 
To remove the  cutoff, we need to take $r\rightarrow \infty$. 
A practical way to describe this limit is to write the 
metric and the couplings as 
\be
\label{gR}
g_{\mu\nu}=\epsilon^{-2}g^{\RR}_{\mu\nu}
\ee
and
\be
\label{phiR}
\phi^I=\phi^I(\phi_\RR,\epsilon)
\ee
where $g_{\mu\nu}^{{R}}$ 
and $\phi^I_\RR$ are the renormalized metric and 
couplings which are kept fixed as we send $\epsilon\to 0$.  
The relation between the bare
couplings $\phi^I$ and the renormalized couplings $\phi^I_\RR$ is obtained by 
integrating the RG-flow 
\be
\label{flowup}
\qquad \qquad \epsilon \,
{\partial \phi^I \over \partial\epsilon} 
\, =  \,- \beta^I(\phi) 
\qquad\qquad 
\phi^I=\phi^I_\RR \qquad \quad {\mbox{at}}\ \epsilon \! = \! 1
\ee
{}Form the supergravity perspective, this procedure for
introducing the renormalized couplings means the following. 
Consider the unique classical supergravity trajectory with asymptotic 
boundary conditions specified by the `bare' fields $(\phi^I,g)$. The 
renormalized fields $(\phi^I_\RR, g_\RR)$ then represent
the values of the scalar field on this trajectory
at some finite value of the scale factor, corresponding to some
fixed RG scale.
As we consider only relevant perturbations, the couplings $\phi^I$ will 
actually go to zero as $\epsilon^{4-\Delta_I}$, as we take the 
limit $\epsilon\to 0$ while keeping $\phi_\RR$ fixed. 
The metric $g_{\mu\nu}$ on the other hand diverges. Hence 
we still find, upon
inserting the expressions (\ref{gR}) and (\ref{phiR}) into the action 
$S$, that the various terms contained in $S_{\rm loc}$  diverge in 
the limit $\epsilon\to 0$. For example, the potential term $U$ will
in general be quartically divergent, while $\Phi$ and $M_{IJ}$ contain
at most quadratic divergences. Even the term $\Gamma$ has a potential
logarithmic divergence that has to be removed in the renormalization procedure.
This can be done by adding appropriate counterterms. The
renormalized effective action $\Gamma_\RR$ is defined by
\be
\Gamma_\RR[\phi_\RR,g_\RR]=\lim_{\epsilon\to 0}
\Gamma\mbox{\raisebox{-2pt}{${}_{\rm \! finite}$}}
[\, \phi(\phi_\RR,\epsilon),\, \epsilon^{-2}g_\RR]
\ee
where $\Gamma_{\rm finite}$ is obtained from $\Gamma$ by 
subtracting its divergent part.

We now would like to show that the action $\Gamma_\RR$ again 
satisfies a similar Callan-Symanzik equation as before, but 
now expressed in terms of the renormalized couplings and metric
\be
\label{loflo2}
{1\over \sqrt{g}}
\Bigl(2 g^{\mu\nu}_\RR{\delta\ \over \delta g_\RR^{\mu\nu}}-
\beta^I_\RR (\phi^I_\RR )
{\delta \ \over\delta \phi_\RR^I}\Bigr)\, \Gamma_\RR\, [ \, \phi_\RR, \, g_\RR \, ]
\; =\ \mbox{\rm local terms} 
\ee
The derivation of this relation is basically a change of variables from
$\phi^I$ to $\phi^I_\RR$.  Beta functions are vector fields in the space
of couplings, and therefore we have
\be
\beta^I
{\delta\ \over \delta\phi^I}=\beta^I_\RR{\delta\ \over\delta\phi^I_\RR}
\ee 
If we substitute this and also the following definition of the renormalized
operators 
\be
{\cal O}^{{}^{\rm R}}_I\, = \, {\cal O}_J \, {\partial\phi^J\over
\partial \phi^I_\RR}
\ee
it is straightforward to recover the 
Callan-Symanzik equations for all renormalized $n$-point functions:
\be
\label{CalSymR}
\Bigl(a{\partial\ \over \partial a}+\beta_\RR^I 
\partial_I\Bigr){\large\langle}
{\cal O}^{{}^{\rm R}}_{I_1}(x_1)\ldots {\cal O}^{{}^{\rm R}}_{I_n}(x_n)
{\large\rangle}
+
\sum_{i=1}^n(\gamma_\RR)\mbox{\raisebox{-2pt}{${}_{I_i}{}^{\!\! J_i}$}}\,
\langle{\cal O}^{{}^{\rm R}}_{I_1}(x_1)..{\cal O}^{{}^{\rm R}}_{J_i}(x_i).. {\cal O}_{I_n}^{{}^{\rm R}}(x_n)\rangle\, = \, 0 .
\ee

\section{Conformal anomaly}

In order to recover the conformal anomaly, let us relax the condition 
that the metric is flat, and use the constraint
(\ref{ham1}) to compute an expression for the trace of the finite part of the 
expectation value of the stress tensor:
\be
\langle T_{\mu\nu}\rangle= 
{1\over \sqrt{g}}{\delta\Gamma\over \delta g^{\mu\nu}}
\ee
The calculation is again similar as outlined above.
We find that the trace anomaly takes form
\be
\label{cs}
2 \langle \, T \, \rangle \, = \, \beta^I \partial_I \Gamma \, +\, c \,
R^{\mu\nu} R_{\mu\nu} + d\, R^2  
\ee 
where $T = T^\mu_\mu$, and the curvature squared terms simply arise due 
to the square of the Einstein term in the action $S_{\rm loc}$. 
We recognize this equation as a standard type
expression as dictated by the broken scale invariance of the effective 
action $\Gamma$ and the trace anomaly relation. 
The coefficients $c$ and $
d$ are given in terms of $\Omega$ and $\Phi$ by
\be 
c \, =\, -{6\Phi^2 \over U}, \qquad \quad d = {2\over U}
\Bigl( \Phi^2 \!- {3\over 2}\partial_I \Phi\, G^{IJ}\!  \partial_J \Phi\Bigr).
\ee 
We can perform a quantitative check on these coefficients, by going to
the fixed point situation, where all $\del_I$ derivative vanish. A
simple calculation shows we can then express $c$ and $d$ in terms of
the 5-d potential term $V$ as $c = 3d =
(V/ 3)^{-3/2}$, 
which reproduces the expression of the holographic Weyl
anomaly obtained in \cite{kostas}. 
Inclusion of the scalars also leads to good agreement with known 
results \cite{fukuma1}.
In addition, there are some harmless ambiguities in the conformal
anomaly obtained in this way \cite{fukuma2}.
The function $c$ is the analogue of the central charge of the 4-d QFT,
and has been proposed as a candidate $C$-function in
\cite{porratietal} and \cite{nicketal}.

\section{Cosmological constant problem}

In \cite{cosm1,cosm2} techniques similar to the ones described here
were used to shed some new light on the cosmological constant problem.
The idea is to view a compactification of string theory also from
a holographic point of view, in which physics at a particular energy
scale in our world should be viewed as taking place somewhere along
an extra coordinate, the analogue of the radial coordinate $r$. This
coordinate indicates the energy scale. Consider some compactification
of string theory, which typically can be warped, and slice it in two
at some value $r=r_0$. The region $r<r_0$ represents the IR physics,
the region $r>r_0$ the UV physics. Given values of the 4d fields
on this slice, their effective action is given by the sum of the IR
and UV effective actions, 
\be S_{\rm  tot} = S_{\rm IR}[\phi,g] + S_{\rm UV}[\phi,g]
\ee
where both $S_{\rm IR}$ and $S_{\rm UV}$ satisfy the 
Hamilton-Jacobi equation. In addition, the complete solution
should be smooth, so derivatives should not jump at the
location of the slice. This implies that
\be
\frac{\partial S_{\rm tot}}{\partial \phi} =\frac{\partial S_{\rm tot}}{\partial g}=0.
\ee
The argument of \cite{cosm1,cosm2} roughly boils down to the
following. The total action $S_{\rm tot}$ is invariant under moving
the slice along the equations of motion. Therefore, the total vacuum
energy, once it is zero at some energy scale (e.g. because 
of supersymmetry) will remain zero at
other energy scales. As we move the slice around, the vacuum energy
obtained from integrating out degrees of freedom is apparently canceled
by the kinetic energy in the warp factor. 
As further evidence, note
 that we can in particular choose flat slices, so that
the effective cosmological constant on the brane appears to be zero.
 
Some comments are in order.\\
\noindent
-One has to careful in studying warped compactifications this way,
it would be better to treat the whole warped compactification all at once.\\
\noindent
-if one wants to break supersymmetry dynamically in such a scenario,
one presumably needs strong coupling physics and it is not clear that
this classical analysis will still be valid.\\
\noindent
-if supersymmetry is explicitly broken, one will encounter 
explicit sources at some energy scale, and it is not obvious that
such arguments go through.\\
-the total action is invariant under renormalization group flow.
Usually, an argument of Weinberg \cite{wein} then states that the
dependence of the low-energy effective action on some fields is
trivial, and this cannot be used to argue that the cosmological 
constant is zero. The present case could avoid this argument
if the low-energy theory is non-local (all the way down to zero
energy). Such non-localities seem unavoidable if one wants
to be able to slice the space in an arbitrary way. Perhaps there
are distinguished slicings where the nonlocality  vanishes, which
could be a criterion that could help determine precisely which
four-dimensional metric the four-dimensional observer will see
in such compactifications. 

\section{Brane world scenarios}

The techniques described here are also very convenient if one
wants to study brane world scenarios, like the Randall-Sundrum
scenario \cite{rs}. For example, if we consider spaces with a
boundary, we are actually interested in the full action $S$
and not just in $\Gamma$. On such spaces, the graviton wave
function is normalizable, and therefore they are dual to
field theories coupled to gravity, as described by $S$.
The effective action was found to be given by
\be \label{bact}
S[\, \phi\, ,\, g\, ] \, = \Gamma\, [\, \phi\, ,\, g\,] +
\int \!\! \sqrt{g}  \, \Bigl(U(\phi)  +   
\Phi(\phi) R \, +  \frac{1}{2}\partial^\mu\phi^I M_{IJ}(\phi)\, 
\partial_\mu \phi^J \Bigr).
\ee
In (\ref{secondflow}) we gave equations for the couplings
that appear in this expression. Thus we obtain a lot of information
about the potential  $V$ and the effective four-dimensional
Newton  constant $\Phi$ from the Hamilton-Jacobi equations.
{}For AdS-like spaces that are bounded by a brane, the full
action on the brane will be the sum of the brane action
and the action (\ref{bact}).

As another example, we consider a simple model for 
a vanishing cosmological constant without fine tuning
that was described in \cite{kachru,arkani}.
The theory consists of five dimensional gravity,
with zero potential and a single scalar. At $r=0$
there is a four-dimensional brane which represents our
world, with tension $T(\phi)$. For the metric
we take a Poincar\'e invariant metric
\be
ds^2 = dr^2 + e^{2A(r )} \eta_{ij} dx^i dx^j .
\ee
We impose a boundary condition $A$, $\phi$ on
the brane. The total action is the effective action
coming from the region $r<0$, the brane tension, 
and the effective action coming from $r>0$. 
Both effective actions are of the form $\int \sqrt{g} U(\phi)$,
and $U$ satisfies the equation (\ref{fst}), i.e.
\be
\frac{1}{3} U^2 - \frac{1}{2} U'^2 = 0 .
\ee
The general solution for $U$ is therefore
\be
U = c e^{\pm \sqrt{\frac{2}{3}} \phi} .
\ee
{}From the  region $r<0$  we get one function $U_L$,
from the region $r>0$ we get another function  $U_R$.
Hence, the total action reads
\be
S_{\rm tot} = \int \sqrt{g} (U_L(\phi) + T(\phi) + U_R(\phi)) .
\ee
We still have to impose the equations of motion for 
$S_{\rm tot}$. That yields (where $\phi_0$ is the value of
$\phi$ on the brane)
\bea
U_L(\phi_0)  + T(\phi_0) + U_R(\phi_0)& = & 0 \\
U_L'(\phi_0)  + T'(\phi_0) + U_R'(\phi_0)& = & 0 .
\eea
These equations guarantee that the derivatives of fields
have the right  jumps over the brane, also known
as the Israel junction conditions. Here they arise
as a very simple consequence of the equations of 
motion. 

Altogether we have two equations for three unknowns
$c_L,c_R,\phi_0$, and there is always a solution.
This seems suggestive of a theory with a vanishing
cosmological constant without fine tuning. However,
these solutions are always singular, and without a good
interpretation of the singularities, or a good description
of the boundary conditions at the singularity, it is not
clear how serious one should take these models. For a recent
discussion, see \cite{nilles}.

\section{Conclusions}

The holographic renormalization group, and the techniques
described here, are very useful tools in studying various
aspects of the AdS/CFT correspondence, brane world
scenarios, etc. We conclude with some questions and
comments.

\begin{itemize}
\item[-] The renormalization group we get is a local renormalization group,
i.e. the scale of the theory is space-time dependent. Is there a physical 
meaning to such scales? Since the metric is an observable, does that mean
that the observed metric and the metric as it appears in the action do
not necessarily have to match each other?
\item[-] We have restricted our attention to the metric and scalar fields
only. It is also possible to include higher form fields and fermions in the
discussion, see \cite{ext}. For fermions some additional subtleties arise
because one cannot choose arbitrary boundary conditions.
\item[-] We have not discussed the boundary conditions in the IR so far.
The Hamilton-Jacobi equations do not completely determine the effective action
$S$. The freedom in $S$ that remains is related to the boundary conditions
in the IR. In the case of the Euclidean AdS/CFT correspondence, we have in mind
that the space is topologically a five-ball, and the IR boundary condition
is such that the fields have a smooth extension over the interior of the
ball. This in principle completely determines the effective action $S$.
It would be very interesting to know how other boundary conditions in 
the IR are reflected in the choice of $S$, especially for confining theories
like \cite{klst,manuc}.
\item[-] Generically, one expects that solutions of the equations of motion
are such that a singularity arises at some finite value of $r$. Criteria for
admissable singularities have been discussed in \cite{gub,manu}, and it
is an interesting question whether there is a relation between admissable
singularities and the behavior of $S$. Perhaps the existence of local 
counterterms that remove the divergences will get a natural interpretation in
this context.
\item[-] So far, the entire discussion has been purely classical. It is difficult
to study quantum corrections, because this requires us to study the full
quantum mechanical equation (\ref{seq}). In this context this equation is known
as the Wheeler-de Witt equation. It is ill-defined as it stands and needs to
be regulated. One thing one can do is to apply it to a subset of theories, e.g.
to the set of Poincar\'e invariant solutions \cite{dbsi}. We already saw that the theory
reduces to quantum mechanics, and the Wheeler-de Witt equation to an ordinary
Schr\"odinger equation. Similar approximations in the case of quantum gravity
are called minisuperspace approximations. One can apply this approximation to
study the quantum stability of some of the brane-world scenario's, but the
validity of such approximations is questionable, since they are not the
leading terms in some systematic expansion. Still, one finds surprisingly enough
that the simple model discussed in the previous section not only solves the
classical Hamilton-Jacobi equation, but also the full Wheeler-de Witt equation
in the minisuperspace approximation. It remains to be seen whether this has
any deeper meaning.
\item[-] An issue raised before is whether there are always local counterterms
that remove the leading singularities from the expansion, or whether in different
circumstances there can be more exotic nonlocal behavior of $S$. This would have
important consequences for the cosmological constant problem, and the analysis
of warped compactifications in general. The application of the techniques in this
paper to more general warped compactifications has not been fully explored,
and it would be interesting to do so. For a discussion of warped compactifications
related to the present discussion see \cite{hv}.
\item[-] Since the flows we consider here have a dual interpretation as
a renormalization group flow on the string world-sheet, one wonders whether
these techniques are useful to study more general renormalization group flows
in string theory, and perhaps have applications to issues like tachyon 
condensation.
\item[-] The Hamilton-Jacobi equations and their quantum extensions look
very similar to Polchinski's version of the exact renormalization group
\cite{Polch}, for a discussion see \cite{lowe}. One can also derive
the renormalization group directly from the renormalization of the string
world-sheet theory \cite{khve}. On the other hand, the renormalization
of string theory is also closely connected to Polchinki's exact renormalization
group and the Batalin-Vilkovisky structure of string field theory \cite{bda}.
Ultimately, one would like to understand all these connections in detail
and to provide a detailed derivation of the renormalization group flow
directly from string field theory. 
\end{itemize}

\end{document}